\input harvmac

\def \ep{\epsilon}
\def\D{\Delta}

\def \k {\kappa} 

\def \g {\gamma}

\def \ha  {1/2}

\def \D {\Delta}
\def \a {\alpha}

\def \p {\phi}

\def \l {\lambda}
\def \t {\theta}
\def \td {\tilde }

\def \inv {^{-1}}
\def \ov {\over }

\def \D {\Delta}

\def \lr { \lref}
\def\np {{  Nucl. Phys. }}
\def \pl {{  Phys. Lett. }}

\def \pr  {{ Phys. Rev. }}

\def \ijmp {{ Int. J. Mod. Phys. }}

\baselineskip8pt
\Title{
\vbox
{\baselineskip 6pt{\hbox{ }}{\hbox
{Imperial/TP/96-97/53}}{\hbox{hep-th/9707134}} {\hbox{
  }}} }
{\vbox{\centerline {One-loop  four-graviton  amplitude }
\vskip4pt
 \centerline {in  eleven-dimensional supergravity }
}}
\vskip -27 true pt
\centerline  {  J.G. Russo{\footnote {$^*$} {e-mail address:
j.russo@ic.ac.uk
 } } and 
 A.A. Tseytlin\footnote{$^{\star}$}{\baselineskip8pt
e-mail address: tseytlin@ic.ac.uk}\footnote{$^{\dagger}$}
{\baselineskip8pt  Also at  Lebedev  Physics
Institute, Moscow.} 
}

\bigskip

 \centerline {\it  Theoretical Physics Group, Blackett Laboratory,}
\smallskip
\centerline {\it  Imperial College,  London SW7 2BZ, U.K. }

\medskip\bigskip

\centerline {\bf Abstract}
\medskip
\baselineskip10pt
\noindent
We find explicit expression for the one-loop four-graviton amplitude in eleven-dimensional supergravity compactified on a circle.
Represented in terms of the string coupling (proportional to the 
compactification radius) it takes the form of an infinite sum 
of perturbative string loop corrections. We also compute the  
amplitude in the case of compactification on a 2-torus, 
which is given by 
an $SL(2,{\bf Z})$ invariant expansion in powers of the torus area.
We discuss the structure of quantum corrections in 
eleven-dimensional theory and their 
relation to string theory.

\medskip
\Date {July 1997}
\noblackbox
\baselineskip 14pt plus 2pt minus 2pt
\lr\mina{M.J. Duff, J.T. Liu and R. Minasian, 
\np B452 (1995) 261, hep-th/9506126.}

\lr\town{P.K. Townsend, hep-th/9512062.}
\lr\kap{D. Kaplan and J. Michelson, hep-th/9510053.}
\lr\hult{
C.M. Hull and P.K. Townsend, Nucl. Phys. { B438} (1995) 109;
P.K. Townsend, Phys. Lett. {B350} (1995) 184;
E. Witten, \np B443 (1995) 85; 
J.H. Schwarz,  \pl B367 (1996) 97, hep-th/9510086, hep-th/9601077;
P.K. Townsend, hep-th/9507048;
M.J. Duff, J.T. Liu and R. Minasian, 
\np B452 (1995) 261, hep-th/9506126.}

\lr \green{M.B. Green and M. Gutperle, hep-th/9604091.}

\lr \berg{E. Bergshoeff, C. Hull and T. Ort\' \i n, \np B451 (1995) 547, hep-th/9504081.}

\lr \gibbon{G.W. Gibbons, \np B207 (1982) 337. }
\lr \hullo{C.M. Hull, \pl B139 (1984) 39. }
\lr \horts{G.T. Horowitz and A.A.  Tseytlin, \pr D51 (1994) 3351,
hep-th/9408040.}

\lr \townelev{ P.K. Townsend, \pl B350 (1995) 184, hep-th/9501068.  }
\lr \calmalpeet {C.G. Callan, J.M.  Maldacena and A.W. Peet, 
\np B475 (1996) 645,  hep-th/9510134. }
\lr \witten {E. Witten, \np B460 (1995) 335, hep-th/9510135.}

\lr \tow {P.K. Townsend,  hep-th/9609217. } 

\lr \schwa  {J.H. Schwarz,
hep-th/9607201.  }
\lr \duff { M.J. Duff, hep-th/9608117.  }
\lr \bergsh{ E. Bergshoeff, E.  Sezgin and P.K. Townsend, \pl B189 (1987)
75.}

\lr\banks{
T. Banks, W. Fischler, S.H. Shenker and L. Susskind, 
   hep-th/9610043.}
\lr \tser{A.A. Tseytlin, hep-th/9609212. }
\lr\russo {J.G. Russo, 
hep-th/9610018 .}
\lr \schwar{J.H. Schwarz, \pl B367 (1996) 97.}
\lr \gsh {M.B. Green  and J.H. Schwarz, \np B198 (1982) 441.}
\lr \gsb {
 M.B. Green, J.H. Schwarz and L. Brink,  \np B198 (1982) 474. }
\lr \mets { R.R. Metsaev and A.A.  Tseytlin, \np B298 (1988) 109. }
\lr \frats {E.S. Fradkin   and A.A. Tseytlin, \np B227 (1983) 252.}
\lr \duftom{ M.J. Duff and  D.J. Toms, 
in:  {\it Unification of the Fundamental 
Particle Interactions II},  Proceedings of the Europhysics Study Conference, 
Erice, 4-16 October 1981, 
ed. by J. Ellis and S. Ferrara (Plenum, 1983).} 
\lr \kallosh { R. Kallosh,  unpublished; P. Howe, unpublished. }
\lr \gris {M. Grisaru and W. Siegel,  \np B201 (1982) 292.}
\lr \sund {B. Sundborg,  \np B306 (1988) 545.}
\lr \acv { D. Amati, M. Ciafaloni and G. Veneziano, \ijmp  3 (1988) 615;
 \np B347 (1990) 550.}
\lr \lip { L.N. Lipatov, \pl B116 (1982) 411; \np B365 (1991) 614.}
\lr \cia {  M. Ademollo, A. Bellini and M. Ciafaloni, \np B393 (1993) 79.}
\lr \grisa{M.T. Grisaru, A.E.M. van de Ven and D. Zanon, \np B277 (1986) 388, 409.}
\lr \grow{D.J. Gross and E. Witten, \np B277 (1986) 1.}
\lr \ban{ T. Banks, W. 
Fischler, S.H. Shenker and L. Susskind,  \pr  {D55}  (1997) 5112,
hep-th/9610043.}
\lr \bbpt{ K. Becker, M. Becker, J. Polchinski and A.A. Tseytlin, 
hep-th/9706072.   } 
\lr \sak { N. Sakai and Y. Tanii, \np B287 (1987) 457.}
\lr \town { Townsend; }
\lr \witt { E. Witten, \np B443 (1995) 85, hep-th/9503124.}
\lr \ggv { M.B. Green, M.  Gutperle and P.  Vanhove, hep-th/9706175. }

\lr \polch {J. Polchinski, {\it  TASI Lectures on D-branes},
hep-th/9611050.}
\lr \ele{E. Cremmer, B. Julia and J. Scherk, \pl B76 (1978) 409.} 
\lr \gv {M.B. Green and P. Vanhove, hep-th/9704145.}

\lr \grgu{M.B. Green and M. Gutperle, hep-th/9701093.}
\lr \tte{A.A. Tseytlin, \np B467 (1996) 383, hep-th/9512081.}
\lr \ber{ M. Bershadsky, S. Cecotti, H. Ooguri and C. Vafa, Commun. Math. Phys.
165 (1994) 31,  hep-th/9309140.}
\lr \anton{
 I. Antoniadis, E. Gava, K.S. Narain and T.R. Taylor, \np B455 (1995) 109,
hep-th/9507115.}
\lr \kir{E. Kiritsis and B. Pioline, hep-th/9707018.}

\lr \zan{ M.T.  Grisaru and D.  Zanon,
\pl B177 (1986) 347; M.D. Freeman, C.N. Pope, M.F. Sohnius and K.S. Stelle, \pl
B178 (1986) 199; Q.-H. Park and D. Zanon, \pr D35 (1987) 4038.  }

\lr \vaw{C. Vafa and E. Witten, \np B447  (1995) 545, hep-th/9505053 
.}

\lr\roo {M. de Roo, H. Suelmann and A. Wiedemann, \pl B280 (1992) 39; \np B405
(1993) 326;  H. Suelmann, ``Supersymmetry and string effective actions", Ph.D.
thesis, Groningen, 1994. }

\def\aa{ {\cal V } }

\def \LL {{\cal H}}
\def \z { \zeta}
\def \k { \kappa_{11}}
\def \R  { R_{11}} 

\def \D {$D=11$\ }
\def \l {l_{11}}
\def \L { \Lambda_{11}}
\def \RR {{\cal R}}
\def \t {\tau}

\def \rr {  R_{11}} 

\def \O {\Omega}

\newsec{Introduction}

Recent suggestions   
indicate that  $D=11$ supergravity 
is a  low-energy effective field theory of  
a more fundamental  M-theory  \refs{\townelev,\witt}
(for  reviews  see  \refs{\schwa,\duff,\tow}).
 One expects that various properties of 
ten-dimensional  string theories 
may  be understood from   eleven-dimensional perspective. 

Most of  known   relations  between  type IIA string theory  and M-theory, 
viewed as its strong-coupling limit, 
 are  restricted to BPS states. 
A surprising recent observation \ggv\ 
  is  that the  {\it tree-level}   type II 
string   correction $ \z (3)  \a'^3   \RR^4$   \refs{\grisa,\grow}
can be interpreted   as originating from  a {\it one-loop}
$D=11$ supergravity contribution. 
Our aim below will be to  compute  the  one-loop 
 four-graviton amplitude 
 in $D=11$ supergravity compactified on a circle   and to demonstrate
 that it   has   the  structure of  an infinite  sum of 
 perturbative  higher-loop string corrections.
This  suggests that the one-loop  quantum $D=11$ theory
 (with a proper
UV cutoff implied by string theory) 
may  contain information  about  certain   string corrections
to  all orders in  string coupling.

The reason why the $D=11$ amplitude has this form may be understood as  follows.
 The one-loop contribution   to  the effective 
action  of  $D=11$ 
supergravity  compactified on a circle of radius $\R$
can be represented as  the  one-loop
correction in  type IIA $D=10$  supergravity plus  an   infinite sum
 of  one-loop  contributions  of massive Kaluza-Klein  modes  
(0-brane supermultiplets).  That sum  may  be represented 
 as a  local series  using  inverse mass expansion, $\sum M^{-2n}  C_n $. 
Since the masses of Kaluza-Klein modes are proportional 
to inverse string coupling 
\witt,  $M \sim \R\inv \sim g_s\inv$, 
the contribution of Kaluza-Klein modes  
has the  structure of  a sum of  perturbative higher-loop 
  closed string corrections,  $\sum_n g_s^{2n} C'_n$.
This suggests that some  perturbative string theory results 
may  be reproduced in the `dual' formulation of the 
 theory,  in which  certain  solitons
(0-branes)   play a   central   role.

The scattering amplitude computed below  corresponds to external gravitons 
with vanishing values of the 11-th component of momentum $p_{11}$.
 Using  
$D=11$ Lorentz invariance  it  is, in principle, 
straightforward   to generalise
the  final   expression for the  amplitude 
 to the case when  external momenta 
are arbitrary, subject only to  the zero-mass  on-shell condition in $D=11$.
The resulting amplitude with $p_{11} = $ fixed 
may then be  interpreted as a 
  one-loop correction to the scattering of 0-branes 
in $D=10$  and   may be of interest from the point of view 
of testing Matrix theory \refs{\ban}.
In  particular, one   should  be able to analise the one-loop $D=11$
supergravity 
contribution  to the phase shift, which was   previously obtained
only 
in  a  semiclassical (eikonal) 
approximation (see \bbpt\ and refs. there).

In Section 2 we shall make some general  remarks 
on cutoff dependence of the $D=11$ supergravity effective action, 
suggesting that certain curvature invariants  should play 
 a special role  in  both $D=11$ and $D=10$ theories.  
 The  
one-loop four-graviton amplitude in $D=11$ supergravity 
 on a circle 
will be  computed explicitly 
in Section 3.1. The  amplitude  in  the supergravity compactified 
 on a 2-torus 
 will be  found in Section 3.2.  
In Section 4 we shall discuss  possible 
 relation of these  amplitudes to perturbative and non-perturbative  
contributions in  string theory.

\newsec{Higher order corrections   in  $D=11$ theory  and 
their relation to string theory}

 Let us  start  with some comments on the structure of higher-loop  terms  in  low-energy $D=11$ supergravity  effective action and their relation to string theory.
We shall  consider the \D theory compactified on a circle  of radius $\R$
with 
the  action 
\eqn\acc{
S =  - {1 \ov 2 \k^2 } \int d^{11} x \sqrt{-g }\ \RR + ...\  ,   \ \ 
\ \ \ \ \   \k^2 = 16 \pi^5 \l^9\ , }
 where $\l$ is the  $D=11$  Planck  
scale.  The two parameters  of  the  compactified 
$D=11$ theory $\R$ and $\k$ are 
related to the string  scale $l_{10}= \sqrt {\a'}$ and the  string coupling  $g_s$ (defined
as a ratio of the fundamental string and  D-string 
tensions \polch) by 
\eqn\rela{\l= (2\pi g_s)^{1/3} \sqrt {\a'}\ , \ \ \ 
 \R = g_s \sqrt {\a'}\ ,  \ \ \ \kappa^2_{10} =   { \k^2 \ov  2\pi \R}
 = 64 \pi^7  g_s^2 \a'^4\ ,    }
 $$\a'= {\l^3 \ov  2\pi \R} , \ \  \ \ \ \ \  \ \ 
 g^2_s =  {2 \pi \R^3 \ov \l^3} \  . $$
The $D=11$ supergravity is  UV divergent, so  one needs to  introduce a
cutoff  $\L$.
Since
 the  $D=11$ and $D=10$ supergravities are related 
by dimensional reduction, $\L$  should be proportional  to 
a cutoff $\Lambda_{10}$ 
in   type IIA  $D=10$ supergravity. 
The two cutoffs  may be related, e.g., 
by comparing  the divergent terms in the 
 one-loop effective actions in $D=11$ and $D=10$  supergravities.
The $D=10$  supergravity is a low-energy 
limit of type IIA string theory, so   
its effective cutoff is $\Lambda_{10} \sim { 1 \ov \sqrt{\a'}}$.  
Expressed in terms of 
the (proper-time) cutoff  $\Lambda_{10}$,   the cutoff $\L$ is  given by 
\eqn\too{   \R  \L^3 =  a   \Lambda_{10}^2\ , \ \ \ \ \ \  \ 
\Lambda_{10}^2 = {1 \ov  2\pi \a'  } \ , }
 where $a$ is a   numerical constant. 
%
%
Eq. \rela\ implies that 
\eqn\cutt{
\L =   a^{1/3} \l^{-1}  \sim   \k^{-2/9} \ , }
i.e.  that $\L$    depends
only on $\k$  and not on $\R$.
This  has a natural `membrane-theory' interpretation:
just as  the  $D=10$  cutoff $\Lambda_{10}$  is  proportional to 
 the square root of the string tension
$T_1 = {1 \ov  2\pi \a'  } = {1 \ov  2\pi l_{10}^2  }$, 
the $D=11$ cutoff $\L$  is proportional to the 
cubic root of the membrane tension 
\eqn\mem {\L =   (2 \pi a T_2)^{1/3}   \ , \ \ \ \ \ \ \  
 T_2 = { 1 \ov  2\pi l_{11}^3  }  = { 1 \ov  4\pi^2 g_s \a'^{3/2}  }   \ . }

The general structure of the cutoff-dependent  part
 of the  effective  action    of  $D=11$ supergravity
at  the  $L$-loop level is
\eqn\stru{
S_{L \infty}  = \k^{2(L-1)}  \sum  
 \L^{n} (\ln \L)^l  \int d^{11} x \sqrt {-g} \ \RR^{m}  \ ,   
    }
where $\RR^m$ stands for  all possible scalars
built out of curvature and its covariant derivatives 
 which have length dimension $-2m$.\foot{We  shall ignore terms 
depending on  3-form field $C_3$ and gravitino.
The structure of terms
depending on $C_3$ is restricted by the  invariance  of the supergravity action 
\ele\
under $C_3 \to -C_3, \  t \to - t$ \duftom.} 
On dimensional grounds,
\eqn\sys{
  n+2m = 9 (L-1) + 11 \  .  
}
Note that purely logarithmic  divergences   ($n=0$)  may  appear only at 
even loop orders and  have  $m=10,19,...$.

At the  one-loop order, the  leading $\RR^m$ ($m=0,1,2,3$)   divergences
 cancel out \frats, so that 
\eqn\div{
S_{ 1  \infty}  \propto 
 \int d^{11} x \sqrt {-g}\    \L^{3} \RR^{4}  \ . 
}
The presence of the cubic $\RR^4$ divergence in $D=11$ supergravity
is implied 
\refs{\frats} \ by the 
presence of quadratic $\RR^4$ divergence in the  $D=10$ supergravity, 
which, in turn,  can be  found 
 as the  $\a'\to 0$ limit  \refs{\gsb,\mets}   of   the  
 one-loop  string-theory 
contribution $ {1\ov \a'} \int d^{10} x \sqrt {-g} \RR^4$  
\refs{\sak,\gsh}.

Eq.\div\ may, in principle, contain also  a linear divergence $  \L \RR^5 $
which  would 
correspond to  the  logarithmic divergence in $D=10$ supergravity
or to a  finite  one-loop term 
$\ln  \a'\ \RR^5$  in the string theory effective  action. Such  $\RR^5$ terms  should   be built out of 
five  powers of the curvature: terms like  $\nabla^2 \RR^4$
 are  absent since
the string  theory {\it four}-graviton amplitude  does not contain 
the corresponding  (momentum)$^{10}$ term \mets .

An  uncompactified $D=11$  M-theory  (having $D=11$ supergravity as its low-energy approximation)  is suggested  to be   a strong-coupling 
limit of type IIA string theory \witt. 
Let us suppose that there are  special  terms 
 $f(g_s) \RR^m$ in the string theory effective action  which 
do not receive corrections beyond certain   order  $L$ in string loop expansion.
Then their coefficients  will have  simple power-like (or `perturbative') 
dependence on $g_s$ in the limit of $g_s \gg 1$, i.e. $f(g_s) \sim g_s^{2(L-1)}$.
Such terms must then have a  natural $D=11$ theory interpretation.
Using this logic, 
one  may be able to  
obtain  certain constraints on  possible terms  in 
the effective action of M-theory. 
As  we   will 
argue  below,  such  special terms in the  string-theory action  may 
correspond to  covariant 
  $\RR^m$  terms in the uncompactified $D=11$ 
 theory  only if $m=3k+1$,\ \ $k=0,1,2,...$.

Using  \rela,\cutt
$$
\int d^{11} x \to 2\pi \R \int d^{10} x \ , \ \ \ \ 
\k^2\sim g_s^{3}\ ,\  \ \ \ \ \R \sim g_s\ ,\ \ \   \L \sim g_s^{-1/3}\ , 
$$
and $ds^2_{11}=dx^2_{11}+ g_{\mu\nu}dx^\mu dx^\nu $ one finds 
$$  
\k^{2(L-1)} \L^n  \int d^{11} x \sqrt {-g}\ \RR^m\ \ \to \ \ g_s^{{2\ov 3}( m-4)} 
 \int d^{10} x \sqrt {-g}\ \RR^m\ .
$$ 
In the last relation we have used \sys.
The condition ${1 \ov 3}(m-4)=k-1$ 
 where $k$ is an integer (effective loop order 
in string theory) implies\foot{The same condition  is found by 
demanding that the dilaton dependence of the $\sqrt {-g} \RR^m$ term 
after the reduction to $D=10$ 
should be $e^{2(k-1)\p}$. Indeed, 
relating the $D=11$ metric
 to the $D=10$  string-frame metric 
by 
$ds^2_{11} = e^{4\p/3} dx_{11}^2 + e^{-2\p/3} ds^2_{10}$
we find that $ (\sqrt {-g} \ \RR^{m})_{11} $
reduces to $  e^{2 (m-4)\p/3 } (\sqrt {-g} \ \RR^{m})_{10}$
so that 
the required condition is  $m-4 = 3(k -1) $ or $m= 3k +1$.}
\eqn\ipi{
m= 3k + 1 \ , \ \ \ \ \  n= 9L -6k \ , \ \ \ \   \ k=0,1,2,... \ . }
 Thus  the  terms in the  $D=11$ action  related to  the  special 
string-theory terms  with coefficients which  have  `perturbative' 
dependence on $g_s \gg 1$  are 
\eqn\thu{
 \k^{2(L-1)}  \L^{9L-6k} \int d^{11}x \sqrt {-g}\ \RR^{3k+1}\ 
  \sim \    l_{11}^{6k-9} \int d^{11}x \sqrt {-g}\ \RR^{3k+1}
 \ ,  }
where we have used \cutt.\foot{
Let us note that supersymmetry may also impose
certain constraints on possible $\RR^{m}$ curvature invariants.
The $\RR^m$ invariants that 
originate from  the   {\it  full} (on-shell) 
superspace integral \refs{\kallosh,\gris}, 
$\int d^{11} x\  d^{32} \theta\  {\cal D}^{2p }\ W^m ,$
\ \  $ W \sim \theta^2 \RR + ...$, 
have   $m=16 + p $ \ 
(combined with \sys\ with $n=0$,  this gives 
further restriction on possible  purely-logarithmic counterterms:
 $ m + p=  9L -14$\  \duftom ). This  condition
includes $m=3k+1 \geq 16$  for $p=3k-15$.
 The terms  with  $m=3k+1  < 16$  \  
(i.e. $\RR^4, \ \RR^7$, etc.)  should  correspond
 to  super-invariants  constructed  as    integrals 
over  parts of  superspace. }

One  may arrive at the same  restriction on powers of curvature invariants 
in the   $D=11$ theory (i.e. $m=1,4,7,10,...$) by    
an    independent argument.
 In general, local perturbative 
 contributions to  the string-theory effective action
are given by  series of terms in expansion in   string 
coupling  and inverse string tension, 
$ \sum \kappa_{10}^{2(L-1)} T_1^{-n} \int d^{10}x \sqrt {-g} \RR^m$, where 
on dimensional grounds,  $m = 2(L-1) + n +  5$.
The  natural  parameter  in M-theory has dimension
(length)$^{-3}$, which  may  be interpreted as the   membrane tension $T_2$.
If we  assume that  the M-theory effective action should  similarly 
contain only terms which  may appear in 
 expansion   in integer powers of  inverse membrane tension,
then the only possible  curvature 
invariants  will be  those given in eq. \thu .
Indeed, 
$ T_2^{-n} \int d^{11}x \sqrt {-g} \RR^m  \sim  l_{11}^{3n}
 \int d^{11}x \sqrt {-g} \RR^m$, so that $2m= 3n + 11$. Since $m$ is a
positive integer,  $n$  must be  an odd number of the form  $n=2k-3$, 
$k=0,1,2,...$, and hence $m=3k+1$.

To summarise, a term $f(g_s)\RR^{3k+1}$ in 
type IIA superstring theory
 corresponds to a covariant term in the eleven-dimensional
 Lagrangian  only 
if it scales  like $g_s^{2k-2}$  in the limit $g_s^2\gg 1$.
Although it is not excluded that  the sum of  an infinite number of 
string  loop corrections 
may  behave like $g_s^{2k-2}$ at strong coupling,  
the non-renormalization of the  $\RR^{3k+1}$  terms 
seems  a natural generalization of the 
suggestion about the 
non-renormalisation of $\RR^4$ term made in \refs{\ggv,\gv} 
to the case of $k>1$.
Thus  we conjecture 
 that all $\RR^{3k+1}$ 
terms   should  not receive contributions
beyond  the $k$-th loop  order in  
type IIA string perturbation theory.\foot{The existence of 
terms  in uncompactified type II string theory 
action  which receive corrections only at  one specific loop order
was conjectured   in \tte.  Examples of such terms
are known in the case of $N=2, D=4$  supersymmetric compactifications 
of type II string theory  \refs{\ber,\anton}.}

At the same time, 
contributions  to $\RR^{3k+1}$  terms
 at   {\it lower} loop  orders in   string perturbation theory 
   are  not excluded (as they will be subleading 
in the $g_s \to \infty$ limit).
Their  $D=11$ 
origin  should be 
 in  the  finite  `Casimir-type'  $\R^{-n}$ terms, which accompany
$\L^n$-terms  when the  $D=11$ effective action is computed 
in  the   space with one circular dimension.
For example, 
  the one-loop $ \L^3 \RR^4$ term 
  in the case of finite radius  $\R$
is  replaced by 
$( \L^3 + c_1 \R^{-3}) \RR^4$ \ \ggv.
In general,\foot{
It may seem   that in  the compactified  case  one should have 
$\L^n \to 
\L^n + a_1 {\L^{n-1} \ov R_{11}}  + ... + a_n {1 \ov R^n_{11}}.$
However, the presence of 
${\L^{n-k} \ov R^k_{11}} $ terms  is  ruled out  on the grounds of locality 
of  UV divergences.}
$$
  \k^{2(L-1)} \R^{-n}   \int d^{11} x \sqrt {-g}\ \RR^m \  \ 
 \to \   \ g_s^{2q} 
 \int d^{10} x \sqrt \RR^m \ ,\ \ \ \  
q= m-3L- 2\ ,
$$ 
where we have used  \sys. 
Remarkably, if  $m=3L+1$ as in \ipi,\thu,   then $q= -1$, i.e.
we conclude that the term 
$  \k^{2(L-1)} (\L^n + c \R^{-n} )  \int d^{11} x \sqrt {-g} \RR^m $
in  the $D=11$ effective action corresponds
to  a   sum of   $L$-loop and tree-level $\RR^{3L+1} $
terms in the  $D=10$ string theory effective action. 
For example,  like the one-loop  $D=11$ terms 
$( \L^3 + c_1 \R^{-3}) \RR^4$, which   correspond to a sum of 
one-loop and tree-level terms in $D=10$ \ggv, 
the two-loop terms 
$\k^2 ( \L^6 + {c_2 \R^{-6}}) \RR^7$ 
should correspond to a sum of 
two-loop ($ { \kappa^2_{10} \ov \a'^2}  \RR^7$) 
 and tree-level  ($ {  \a'^6 \ov \kappa^2_{10}}  \RR^7$)  
terms in   string theory.

\newsec{ One-loop  four-graviton amplitude in $D=11$ supergravity}

Deriving  the one-loop  four-graviton amplitude directly from the component 
formulation  of $D=11$ supergravity \ele\   would be quite  complicated.
Fortunately, there is a short-cut way 
using  the  known   expression  \gsh\  for the 
one-loop   $D=10$ closed  superstring  
4-point amplitude. 
It was shown in  \gsb\ that the  one-loop 
 graviton scattering amplitude
 in  $D \le 8$ maximal supergravities 
 can  be obtained as a limit   
($\a'\to 0, R\to 0, \kappa_D=$fixed)   of the   amplitude 
of  $D=10$ string theory
compactified on a torus. 
To find the amplitude in $D=10$ type II supergravity theory 
one should  take $\a'\to 0$ limit treating $1/\a'$ as a proper-time
 UV cutoff \mets.
The resulting expression is  formally the same as  for $D <8$ \gsb\
(where the  $\a'\to 0$  limit is regular),   but it
  still depends on $\a'$
via  the cutoff (and it is quadratically divergent for  $\a'\to 0$).

According to  \gsb\  
$$
 A_4^{(D)} =  \left( {2\pi R \ov \sqrt{\a'}}\right) ^{D-10}\ 
{\kappa_{10}^2 \ov \a'} 
  \int^\infty_1 d\t_2\ \t_2^{6-D \ov 2}  \ F(s,t;\t_2)   \   
$$ 
\eqn\ampli{
= \ 
{\kappa_{D}^2 } 
  \int^\infty_{\a' }
 d\t \ \t^{6-D \ov 2} \   F(s,t;\t) 
\  .
}
Here and in what follows we omit the standard kinematic factor 
$K \sim $ (momentum)$^8$ in the expressions for the  four-graviton amplitude
and ignore the overall normalization coefficient.
In eq. \ampli\ 
 $ {\kappa_{D}^2 } =
 (2\pi R)^{D-10}
{\kappa_{10}^2 }$, \ $\t \equiv \a' \t_2$,  \ 
and 
 \eqn\bbb{
F( s,t;\tau )=\int [d\rho] \  e^{- \tau  M(s,t;\rho )}\ ,\ 
}
$$
\int [d\rho]\   e^{- \tau  M(s,t;\rho )}\equiv
\int_0^1 d\rho_3  \int_0^{\rho_3} d\rho_2 \int_0^{\rho_2}
d\rho_1\ e^{- \tau  M(s,t;\rho )} 
+   {\rm 5\ terms\ that\ symmetrise \ }s,t,u
$$
\eqn\ccc{
M(s,t;\rho )\equiv  s\rho_1 \rho_2 +t \rho_3\rho_2+u \rho_1\rho_3 +
t (\rho_1-\rho_2\big)\ , \ \ \ \ \ \ \   s+t+u=0 \ . 
}
The dependence on the  cutoff 
$\a'\to 0$   disappears  in $D < 8$ 
 (where maximal supergravities are  one-loop finite),
but  remains  in $D=10$ 
\eqn\ddd{
 A_4^{(10)} =    {\kappa_{10}^2 } 
 \int^\infty_{\ep_{10}}
{ d\t\over \t^{2} } \  F(s,t;\t)  \ \ \sim \ \  \Lambda^2_{10}\  +\ 
 {\rm finite \ part }
 \ .  }
 Here  $\t \equiv \a' \t_2 =   { {\rm t} \ov 2\pi} $
is related to
the standard proper-time parameter ${\rm t}$ so 
that 
  the  effective  proper-time cutoff   
is $\Lambda_{10} =  
{ 1 \ov \sqrt { 2 \pi \ep_{10}} } = {1 \ov \sqrt {2\pi \a'}}$.

\subsec{$D=11$ supergravity compactified on a circle}

As  follows from the string-theory derivation in \gsb\
(and  is obvious from the proper-time integral representation of \ddd)
 the amplitude  in 
the case of $D=10$ supergravity compactified on a circle is 
given  essentially by the $D=9$ supergravity expression 
\ampli\ with   only  one
modification:   the  factor of the sum over the Kaluza-Klein modes
$  \sum_m e^{-{\pi \tau m^2\ov R^2_{10} }}$
should be introduced  under the integral over $\t$.

Being a consequence of the underlying supersymmetry, 
 the same correspondence pattern   applies to  the 
 4-point amplitudes  of 
any pair of maximal supergravities
obtained  by  dimensional reduction,  irrespective of  their dimension
and relation to string theory. 
 The  four-graviton amplitude in $D=11$ supergravity compactified  on a circle
(with all external  particles having ten-dimensional
 polarisations  and $p_{11}=0$)
is thus given by  eq. \ddd\   with  an extra Kaluza-Klein
factor, i.e. 
\eqn\zxx{
 A_4^{(11)} = {\kappa_{11}^2 }  (2\pi \R)\inv    A_4 (s,t)  \ ,  
}
\eqn\dddd{ 
 A_4 (s,t) =   \sum_{m=-\infty}^\infty \int_{\ep_{11}}^\infty 
{d\tau \over \tau^{2}}\  e^{-{\pi \tau m^2\ov \rr^2 }}\  F(s,t;\tau)\  , \ \ \ \ \ \ \ep_{11} = \L ^{-2} \ , 
}
where $F$ is defined in \bbb.
Because of the  sum over the Kaluza-Klein modes, the  $\t$-integral
here  has a stronger (cubic instead of quadratic, cf. \ddd) 
divergence,   as appropriate to  the   $D=11$ theory.

 The resulting amplitude  \dddd\  is in agreement with the 
general expression for  the 
 $D=11$ supergravity  four-graviton 
 amplitude  suggested  (on the  basis of a somewhat  different  reasoning) 
in \ggv.
Our aim  below  will be  to study the structure  of 
 this amplitude,  going beyond the 
leading (momentum)$^8$ terms considered in \ggv.

The integrand in the 
amplitude eq. \dddd\ can be expanded in powers of $M$ 
\eqn\zaaa{
A_4(s,t)= \sum_{m=-\infty}^\infty \int_{\ep_{11}}^\infty {d\tau \ov \tau ^2}  
\  e^{-{\pi \tau m^2\ov \rr^2 }}
\int [d\rho ]\  \sum_{k=0}^\infty  { (-1)^k \ov k!} \tau ^kM^k(s,t) \ .
}
Let us   separate the first  ($k=0$) term  $A_4^{\rm (a)}$ in the expansion,
\eqn\zbbb{
A_4(s,t)=A_4^{\rm (a)}+ A_4^{\rm (b)}(s,t)\ ,\ \ \ \  \ \ \ 
A_4^{\rm (a)} = \sum_{m=-\infty}^\infty \int_{\ep _{11} } ^\infty 
{ d\tau  
\ov \tau^{2} } \ e^{-{\pi \tau m^2\ov \rr^2 }} \ . 
}
The term $A_4^{\rm (a)}$  was considered  in \ggv . 
Performing   first the   Poisson resummation,  one obtains 
 \eqn\eef{
A_4^{\rm (a)}  
=\ R_{11} \sum_{w=-\infty}^{\infty} \int_0^{1/\ep_{11}}  d\hat \tau \ \hat \tau^{\ha} \ 
e^{ -\pi  w^2 \rr^2\hat \tau } \ ,\ \ \ \  \ 
\hat\tau\equiv \tau ^{-1} \ . 
}
Splitting  the sum in  \eef\  into 
the $w=0$  
and $w\not=0$  parts
\eqn\opo{
A_4^{\rm (a)}=A_4^{\rm (a0)}+ \tilde A_4^{\rm (a)} \ , 
}
 we find that 
  $\tilde A_4^{\rm (a)}$ is finite,    while $A_4^{\rm (a0)}$
is the   UV divergent  
contribution
\eqn\efg{
A_4^{\rm (a0)}=\rr \int_0^{1/\ep_{11}}  d\hat \tau \ \hat \tau^{\ha}
= {2\ov 3 }\rr  \ep^{-3/2}_{11} = {2\ov 3 }  \rr  \L^3 \ .  
} 
Thus finally 
\eqn\zccc{
A_4^{\rm (a)}={2\ov 3 }  \rr \L^3\ + \  {\zeta(3)\ov \pi \rr ^2 } \ . 
}

The cutoff-independent part $A_4^{\rm (b)}$ in \zbbb\ can be written as
$A_4^{\rm (b)}(s,t)=A_4^{\rm (b0)}(s,t)+\td A_4^{\rm (b)}(s,t)$,
with $A_4^{\rm (b0)}$ representing the $m=0$ contribution
  to  the sum in  \zaaa, i.e. 
\eqn\zeee{
A_4^{\rm (b0)}(s,t)=
\int [d\rho ] \int _0^\infty {d\tau\ov \tau ^2} \left[e^{-\tau M(s,t)} -1+ \tau M(s,t)\right] \ , 
}
\eqn\zfff{
\td A_4^{\rm (b)}(s,t)=  2 \int [d\rho ]\sum_{m=1}^\infty \int_0^\infty 
{d\tau  \ov \tau^{2}}\ e^{-{\pi \tau m^2\ov \rr^2 }}
 \sum_{k=2}^\infty  {(-1)^k \ov k!}\tau ^k M^k(s,t) \ . 
}
In the last expression \zfff\ we have omitted the  term linear in $M$,
which  drops out after integrating
over  $\rho $ and  symmetrising
in $s,t,u$\   since $s+t+u=0$. This  reflects  the absence 
of logarithmic 
divergences in the 4-point  amplitude in $D=10$ supergravity \mets. 
  To compute $A_4^{\rm (b0)}(s,t)$  in  \zeee\ 
we first regularise it by integrating $\tau $
from 0 to  $\tau _0$  and then take the limit $\tau_0 \to \infty$.
As a result, 
 $$
A_4^{\rm (b0)}(s,t)
=\lim _{\tau_0 \to\infty }
\int [d\rho ]\  \bigg[{\tau_{0}^{-1}\big( 1-e^{-{ \tau_0} M} } \big)
+(\gamma -1+\ln {\tau_0} )  M(s,t) 
$$
\eqn\zggg{
+\ 
 M(s,t) \int_{{\tau_0}}^{\infty } {d\tau \ov \tau } e^{- \tau M(s,t)}
+M(s,t)\ln M(s,t) \bigg]\  =\  \LL(s,t) \ , 
}
where\foot{The  integral over $\rho $ in  $\LL (s,t)$ can be performed explicitly,
giving a combination of  logarithmic and
 polylogarithmic functions.}
\eqn\yey{
\LL(s,t) \equiv \int [d\rho ] \ M(s,t;\rho)\ln  M(s,t;\rho)
= s \bar \LL \big( {s \ov t} \big)  \ ,
}
and  we  used again that  terms  linear in  $M$  disappear 
after symmetrisation in $s,t,u$.
The integration over $\tau  $ in \zfff\  then gives 
\eqn\ziii{
\td A_4^{\rm (b) }(s,t)=\sum_{k=2}^\infty {2 (-1)^k\ov \pi^{k-1} 
k(k-1)}
\sum_{m=1}^\infty  m^{2-2k} \ \rr ^{2k-2}  H_k(s,t) \ 
=\sum_{k=2}^\infty c_{k} \rr ^{2k-2} H_k(s,t) \ .
}
Here 
the coefficients $c_k$  are proportional to  values of the 
 Riemann $\zeta$-function
\eqn\coe{
c_{k}={2(-1)^{k}  \ov \pi ^{k-1}k(k-1)}\ \zeta(2 k-2) \ ,  
}
and 
\eqn\iii{
H_k  (s,t)\equiv \int [d\rho]  \ M^k  (s,t;\rho ) =  s^k \bar H_k \big(
{s\ov t}\big)
\ , 
}
where $\bar H_k$ is a polynomial of order $k$.
The  integral similar to \iii\ 
appeared in \gsb, where it was put   into  the form
\eqn\jjj{
H_k  (s,t)=b_k \big[ I_k  (s,t) +
I_k  (t,s)+I_k  (s,u)+I_k  (u,s)+I_k  (t,u)+
I_k  (u,t)\big]  \  , 
}
$$
b_k={\sqrt{\pi}\  \Gamma(k +1)\ov
2^{2k +2}\Gamma (k+5/2)}\ ,\
$$
\eqn\kkk{
  I_k (t,s)=t^{k +1}\int _0^1 dx {(1-x)^{k +1}\ov sx -t (1-x)}
=-{t^k \ov k+2} \ {}_2F_1[1,1,3+k ; 1 + {s \ov t} ] \ .  
}
For integer $k> 0$  the function $I_k$ 
 reduces to a polynomial plus some combination of logarithmic functions. 
The latter cancel out 
in the symmetric combination $H_k(s,t)$ in eq. \jjj. 
 One is left with  
a homogeneous polynomial of degree $k$ in the variables $s,t,u $
(this follows   also from direct computation of the  integrals in  eq. \iii\ 
after expanding the binomial). 

Combining the above expressions \zccc , \zggg , \ziii , we  find 
\eqn\nnn{
A_4(s,t)= {2\ov 3 }  \rr \L^3  +{\zeta (3)\ov \pi \rr ^2}+ s \bar 
\LL \big( {s \ov t}\big) +
\sum_{k=2}^\infty c_{k} \rr^{2k-2}  s^k \bar H_k\big( {s\ov t}\big) \ .
}
 In the case when all 11 dimensions are non-compact, the amplitude is 
given by the same universal expression \ampli\  (with $D=11$
and 11-dimensional cutoff)
\eqn\ded{
 A_4^{(11)}  =    {\kappa_{11}^2 } 
 \int^\infty_{\ep_{11}}
{ d\t\over \t^{5/2} } \  F(s,t;\t)   =
{\kappa_{11}^2 }  \big[ { 2\ov 3  }    \Lambda^3_{11}\ 
 +\ 
{ 4\ov 3 } \sqrt {\pi}   s^{3/2} \bar H_{3/2}  ({s\ov t}) \big] 
\ ,   }
where $\bar H_{3/2}$ is defined as in \iii.
For comparison, 
the corresponding amplitudes \ampli\ in uncompactified $D=10$ \ddd\ 
and $D=9$ supergravities
  have the following explicit form ($\Lambda_{9}=\Lambda_{10}$)
\eqn\dedd{
 A_4^{(10)} =    {\kappa_{10}^2 }  \big[
{ 2\pi }  \Lambda^2_{10}\
 +   \  s \bar {\cal H}  ({s\ov t})  \big]
\ ,   }
\eqn\dedr{
 A_4^{(9)} =    {\kappa_{9}^2 }  \big[  2\sqrt 2 \Lambda_{9}\ 
  - \ 2 \sqrt \pi    \sqrt s \bar {H}_{1/2}   ({s\ov t})  \big]
\ .   }
Thus the third term  in \nnn\ is  the finite part of the 
contribution of the massless $D=10$ supergravity fields.

\subsec{$D=11$ supergravity compactified on 2-torus}
In the case of  compactification on $T^2$, one has (cf. \zxx,\dddd) 
\eqn\zzxx{
 A_{4\rm T}^{(11)} =  {\kappa_{11}^2 }  (4\pi ^2 R_{10}R_{11} )\inv   
 A_{4\rm T}  (s,t)  \ ,  
}
\eqn\taaa{
A_{4\rm T} (s,t)= \sum_{m,n=-\infty}^\infty \int_{\ep_{11}}^\infty {d\tau \over  
\tau^{3/2} } \ e^{-\pi \tau \big( {m^2\ov \rr^2 }+{n^2\ov R_{10}^2 }\big)   }
F(s,t;\t) \ . 
}
As in the circle case, we expand 
the exponential  $e^{- \tau M}$ in $F$ \bbb\ in powers of $M$ 
and separate the $k=0$ term  as in \zbbb, 
\eqn\tbbb{
A_{4\rm T} (s,t)=A_{4\rm T} ^{\rm (a)}+ A_{4\rm T} ^{\rm (b)}(s,t)\ , 
\ \ \ \ \  A_{4\rm T} ^{\rm (a)} = \sum_{m,n=-\infty}^\infty \int_{\ep_{11}}^\infty 
{d\tau \over  \tau^{3/2}}\  e^{-\pi \tau  \big(
 {m^2\ov \rr^2 }+{n^2\ov R_{10}^2 }\big) } \ . 
}
The constant part $A_{4\rm T} ^{\rm (a)}$ is the one considered in \ggv,
and it can be computed   by Poisson resumming in both $m$ and $n$
and integrating over $\tau$,
\eqn\tccc{
A_{4\rm T} ^{\rm (a)} 
= {2\ov 3 }  \aa \L^3 + {\zeta(3) E_{{3/ 2}} (\Omega )\ov  \pi 
\aa ^{\ha}\ } \ ,
}
where 
\eqn\defo{
\aa \equiv R_{10}\rr\ ,\ \ \ \ \  \ \Omega= \Omega_2 
\equiv  {R_{10}\ov \rr}  \ .  
}
$\L $ is the same cutoff as in \efg \ and 
$E_r(\Omega)$ is the generalised 
Eisenstein series,
\eqn\tddd{
E_r(\Omega ) =     \sum _{(p,q)'} { \Omega^r \ov  
(p^2\Omega^2  +q^2 )^{r}}   =       \sum _{(p,q)'} 
(p^2\Omega +q^2 \Omega^{-1} )^{-r} \  , 
}
where the notation $(p,q)'$ means  that  $p$ and $q$ are relatively prime.
As in \grgu,  one can show that for large $\Omega$
\eqn\expa{
E_r(\Omega ) = \O^r +  \g_r   \O^{1-r} + O(e^{-2\pi \O}) \ ,  }
$$ \g_r = { \sqrt \pi\ \Gamma(r-1/2)\ \zeta(2r -1) \ov
\Gamma(r)\ \zeta(2r) }  \ . $$
To  calculate  $A_{4\rm T} ^{\rm (b)}(s,t)$ in \tbbb\  we  decompose it as  
$A_{4\rm T} ^{\rm (b)}(s,t)=A_{4\rm T} ^{\rm (b0)}(s,t)+\td A_{4\rm T} ^{\rm (b)}(s,t)$,
with $A_{4\rm T} ^{\rm (b0)}$ representing the $(m,n)=(0,0)$ contribution,
$$
A_{4\rm T} ^{\rm (b0)}(s,t)=
\int [d\rho ] \int _0^\infty {d\tau\ov \tau ^{3/2}} \left[e^{-\tau M(s,t)} -1+ \tau M(s,t)\right]
$$
\eqn\teee{
= \ -2\sqrt{\pi }\int [d\rho ] \ M^{\ha}(s,t)  = -2\sqrt{\pi } H_{\ha}(s,t)
= -2\sqrt{\pi } s^{1/2} \bar H_{\ha}\big( {s\over t}\big) \ . 
}
For the remaining part $\td A_{4\rm T} ^{\rm (b)}$ we have (cf. \ziii )
$$
\td A_{4\rm T} ^{\rm (b)}(s,t)=   \int [d\rho ]
\sum_{(m,n)\neq (0,0)} \int_0^\infty {d\tau \ov 
\tau^{3/2} } \ e^{- \pi \tau \big( {m^2\ov \rr^2 } + {n^2\ov R_{10} ^2}\big) }
 \sum_{k=2}^\infty  {(-1)^k\ov k!} \tau ^k M^k  
$$
\eqn\tfff{
=\ \sum_{k=2}^\infty { (-1)^k \Gamma (k- \ha) \ov   \pi ^{k-\ha}\ k!}
 \sum_{(m,n)\neq 0 } \left(
 {m^2 \ov \rr^2 }+{n^2\ov R_{10} ^2}\right) ^{\ha-k} \ H_k(s,t) \ , 
}
or, equivalently (cf. \ziii)
\eqn\tjjj{
\td A_{4\rm T} ^{\rm (b)}(s,t)=\sum_{k=2}^\infty 
d_k  \sum_{(p,q)'} \big( p^2\Omega  +q^2 \Omega^{-1} 
 \big) ^{\ha-k}\  \aa ^{k-\ha} H_k(s,t) \  ,  
}
$$ d_{k}  = {2 (-1)^k\  \Gamma (k- \ha)
   \ov \pi ^{k-\ha}\  k! }\ \zeta (2k-1) \ . $$
The total amplitude 
$A_{4\rm T} (s,t)=A_{4\rm T} ^{\rm (a)}+A_{4\rm T} ^{\rm (b0)}+\td A_{4\rm T} ^{\rm (b)}$
in the 2-torus case 
is thus (cf. \nnn )
$$A_{4\rm T} (s,t)= 
   {2\ov 3 } \aa \L^3+ { \zeta (3) E_{{3/2}}(\Omega )\ov \pi \aa ^{\ha} }
$$
\eqn\tmmm{
 -\ 
2\sqrt{\pi }  s^{1/2}  \bar H_{1/2}\big( {s \ov t}\big)  
+ 
\sum_{k=2}^\infty d_{k} E_{k- \ha}(\Omega )\ \aa ^{k-1/2} s^k \bar 
H_k\big( {s \ov t} \big)  \ . 
 }
Written in this form the amplitude is given by an $SL(2,{\bf Z})$ invariant
expansion in powers 
of 
the  torus area $\sim \aa $. 

In  eqs. \nnn\ and \tmmm\  the functions 
$\bar H_k\big( {s \ov t} \big)$ with $k=2,3,...$ 
are polynomials of degree
$k$, and thus  correspond  to local higher derivative terms in the 
one-loop effective action (these are the contributions of the massive
Kaluza-Klein modes).
 The non-local ($D=10$ massless mode)  contributions  originate  
 from  the $\bar {\cal H}$ term  in the circle amplitude case  \nnn\  or 
from $\bar H_{1/2}$ term in the torus amplitude  case  \tmmm.
The latter $\bar H_{1/2}$ term has the meaning of the finite part of the 
amplitude in $D=9$ supergravity \dedr.

\newsec{Remarks on relation to string theory }

Let us now comment on the structure of the amplitudes \nnn\ and \tmmm ,
corresponding to the circular and toroidal compactifications of the $D=11$ supergravity, and their  relation to string theory.
Expressing   \nnn\  in terms of the  string coupling  $g_s$ and  
the string scale $\sqrt { \a'} $ using \rela,\too\ we find
 \eqn\pqr{
A_4(s,t)=  {a\ov 3\pi \a' }  + {\zeta (3)\ov \pi \a'  g_s^2}+ 
s \bar \LL \big({s \ov t}\big) +
\sum_{k=2}^\infty c_{k}  g_s^{2k-2} \a'^{k-1}  s^k\bar  H_k \big({s \ov t}\big)  
 \ . 
}
The  first two constant  terms  in this amplitude 
(multiplied by the kinematic factor) correspond  to the
one-loop and tree-level $\RR^4$ terms in the  type II string effective 
action.\foot{In the notation of \ggv , 
${2\ov 3}\L^3\equiv C={\pi\ov 3} $, corresponding to  
$a=\pi ^2$ in our notation. 
As was argued
in \ggv\ using the first two terms
in the  amplitude  \tmmm\ on the 2-torus, this  value is  implied by 
consistency with   string theory (T-duality invariance of one-loop term 
in type II  theories compactified on a circle).}
  That the  one-loop 
 amplitude in $D=11$ supergravity
 effectively  includes  \ggv\     the 
tree-level $\zeta (3) \RR^4 $  term  of  string theory
may look   miraculous:  while
in string theory  this term is  produced by 
exchanges of  massive string modes, in 
$D=11$  expression it originates from the loop 
of the Kaluza-Klein modes  which are 0-brane solitons  of string theory.
This fact  
is  suggesting  that the uncompactified  type IIA string theory  
(`dual' to $D=11$ theory) 
 may  have a  reformulation 
in terms of solitonic objects.\foot{The relation between
tree-level $\RR^4$ term in 
 type IIA  theory  and one-loop  $\RR^4$ term in $D=11$ theory 
is reminiscent of the relation between  tree-level $F^4$
term in type I theory and one-loop $F^4$ term in
 heterotic theory \tte.}

Let us briefly comment  on the explicit structure of the 
$\RR^4$ terms in  the effective actions of 
type IIA and $D=11$ theories.
In general, one expects  the  $\RR^4$ terms  in 
$D=10$ theory  to be a 
linear combination  of  the $D=10$ terms
$ {\cal J}_1 \equiv  t_8 t_8  R^4 $ 
and ${\cal J}_2 \equiv  {1 \ov  8 } \ep_{10} \ep_{10}  R^4$.\foot{We follow
 the notation of \tte\ 
 up to the sign change $\ep_{10} \ep_{10}  \to - \ep_{10} \ep_{10} $
due to Minkowski signature used here. Thus  here   
$J_0 = t_8 t_8  R^4  + {1 \ov  8 } \ep_{10} \ep_{10}  R^4$.}
 ${\cal J}_2$ is  the  higher-dimensional 
extension  of the  Gauss-Bonnett invariant in 8 dimensions
($\ep_8 \ep_8 \to - { 1\ov 2} \ep_{10} \ep_{10}$). Its  expansion
near flat space ($g_{mn} = \eta_{mn} + h_{mn}$) 
 starts with $h^5$ terms and thus its coefficient 
cannot be determined from consideration  of the on-shell 
4-graviton amplitude only. 
The sigma-model approach implies \refs{\grisa,\zan}
that (up to the usual field redefinition ambiguities) 
the tree-level type II string term  is 
$L_0\sim \zeta(3) J_0$, \ $J_0 = {\cal J}_1 + {\cal J}_2$.
The  structure of the kinematic factor in  the one-loop 
type IIA  4-point amplitude with transverse polarisations and momenta
$ (t_8 + { 1\ov 2} \ep_8) (t_8 + { 1\ov 2}  \ep_8) $
 hints  that the one-loop $\RR^4$ terms in $D=10$ type IIA theory 
should be proportional 
to the  opposite-sign combination
$ {\cal J}_1 - {\cal J}_2$.\foot{This is implied also by 
 the  discussion in 
  \kir.  We are grateful to E. Kiritsis 
for  clarifying   correspondence   on the issue of 
$\RR^4$ terms.} 
Combining  with  the  $B \RR^4$ term  \vaw,   we get 
$ L_{1A} = {\cal J}_1 - {\cal J}_2 + 
b_1  \ep_{10} B [ \tr R^4  - { 1 \ov 4} (\tr R^2)^2]$.
This can be re-written  
(using ${\cal J}_1 =  24[t_8  \tr R^4  - { 1 \ov 4} t_8 (\tr R^2)^2]$) 
as a combination of  the bosonic parts of the 
{ \it three }
$N=1$ super-invariants \roo\  $ I_3= 
t_8 \tr R^4 - {{1\ov 4}} \ep_{10} B \tr R^4$ , 
\  $I_4= t_8 \tr R^2 \tr R^2  - { {1\ov 4}} \ep_{10} B (\tr R^2)^2$
and  $J_0$ \  {\it provided}  $b_1 =-12$. Then
\ $ L_{1A} = -J_0 + 48 (I_3 -{1\ov 4} I_4)$.\foot{Similar  observation 
was made in \gv, where, however, 
the possible presence of ${\cal J}_2$ was ignored
and  thus to  be able to represent  $L_{1A}$ as a combination of 
$I_3$ and $I_4$ 
 a   different value  $b_1= -6$
was   assumed.} 
While the coefficients of $I_3$ and $I_4$ are expected not to be 
renormalised, there is no reason  for  
this  to  be true   for   the coefficient
of $J_0$ \tte\ and thus of the ${\cal J}_2$ term. 
This may  preclude one from  
identifying the  $D=11$ counterpart of this
term  as $  { 1 \ov 24} \ep_{11} \ep_{11} R^4$.

Returning to the discussion of the amplitude \pqr, 
  we  observe  that  not only the  two constant terms but 
also  all {\it  
 momentum-dependent}  terms in  the $D=11$ amplitude  \pqr\ 
  have  
`perturbative' dependence on the type IIA  string coupling.
It  appears as if  the {\it one-loop}
 four-graviton  amplitude in $D=11$  supergravity 
  represents  a sum of certain  perturbative string  corrections, 
containing  contributions  of 
 {\it all orders}  in  the string loop expansion.

It is not clear, however, 
 which   regions of the moduli spaces of  higher genus Riemann
surfaces this  expression is accounting for.
Moreover, while the first two terms in \pqr\  (or $ \RR^4$ terms
in the  type II string  effective action) 
are expected to be unchanged by both $D=11$ supergravity and 
  type IIA string   higher-loop corrections
 \refs{\gv,\ggv}, 
this  may not  
 be true for  other $s,t$-dependent  terms in \pqr.
If this is the case, one  may be   unable  to relate  the 
 $D=10$ and $D=11$ 
expressions  in a simple way.


To  relate the torus amplitude 
\zzxx,\tmmm\ to type IIA and type IIB
string theories compactified on a circle,
it is useful 
to  consider  the corresponding 
 contribution 
to the  effective action of $D=11$ supergravity compactified
on a  2-torus which 
may be written in the following symbolic form (cf. \div)
 $$
S_1 \propto  \int d^{9} x \sqrt {-g}   \bigg[
{2\ov 3 }  \aa \L^3  +
\pi\inv { \zeta (3) E_{{3/2}}(\Omega ) \aa ^{-\ha} } 
$$
\eqn\coio{
+\   h_{1/2} (\nabla^2)^{1/2} + 
 \sum_{k=2}^\infty h_k  E_{k- \ha}(\Omega )\ \aa ^{k-1/2} 
  \nabla^{2k} \bigg] \   \RR^4\   .  
}
 One  may  now relate the $D=11$ metric $g_{mn}$ and the torus
area 
$(2\pi)^2\aa$ and  the modulus $\O$ 
 to the string-frame metrics, couplings and  radii 
of type II string  theories compactified on a 
circle.\foot{Here we shall 
follow \refs{\gv,\ggv} and 
use the standard 
 relations ($\a'=1$): \ $ds^2_{11} = e^{4\p_A/3} dx_{11}^2 +
e^{-2\p_A/3} ( R^2_A dx^2_{10} + ds^2_{9A})$, \ 
$ds^2_{9A} = ds^2_{9B}$, 
\  $R_A = 
 R_{10} R^{1/2}_{11}=R_B\inv$, \  $g_A = e^{\p_A} = R_{11}^{3/2}$, 
    \ 
 $g_B = e^{\p_B} = R_{11} R_{10}\inv$, \  so that 
\ $R_{10} =   g_B^{-1/3}  R_B^{-2/3}, \ \ 
 R_{11} =   g_B^{2/3}  R_B^{-2/3}$, 
\ \ $ \int d^{9} x \sqrt {-g} \RR^4 
 =  R^{-1/2}_{11}  \int d^{9} x (\sqrt {-g} \RR^4)_B $, \
$ \nabla^2  = R_{11} (\nabla^2)_B$, etc.}
In terms of type IIB coupling and compactification radius,
 $\O =R_{10} R_{11}\inv = 
 g\inv_B, \ \  \aa= R_{10} R_{11} =  \a'^{5/3}  g_B^{1/3}  R_B^{-4/3}$, 
so that the  limit of uncompactified  
type IIB theory   corresponds to  $\aa \to 0$ for fixed $\O$  \schwa . 
The   momentum-dependent terms  in \tmmm\ and 
the higher-derivative terms 
in \coio\ disappear in this limit (the third non-local term
is  also subleading as  it does not scale as $R_{B}$, see below).
The remaining  second term  proportional to the 
  Eisenstein  function   
$E_{{3/2}}(\Omega )$  was shown in \grgu\ 
to  contain  not only the tree-level and one-loop contributions but also 
the sum of all  type IIB D-instanton contributions  to 
 the ${\cal R}^4 $ term (similar results  for 
type IIB theory compactified to 8  dimensions were obtained in \kir).
The  limit of non-compact 
type IIA theory is $R_A\to \infty$ for fixed $g_A$, i.e.
 $R_{10} = \O \aa \to \infty $ for 
fixed $R_{11} = (\aa/\O)^{1/2}$. 
In that limit one recovers the   amplitude \pqr\
of the $D=11$ theory compactified on a circle, 
containing  perturbative  
contributions to all orders in string coupling.

In general, eqs. \tmmm\ and \coio\  appear to be describing
  a mixture of perturbative and non-perturbative contributions
in type II string  theories compactified on a circle.
Expressing \coio\ in terms  of type IIB parameters
and using the expansion \expa\  of $E_r(\O)$  for large $\O$ 
($\O$ is large for small $g_B$)
we find
 $$
S_1 \propto  \int d^{9} x  (\sqrt {-g })_B \ R_B   \bigg(
{2\ov 3 } \L^3 R^{-2}_B
  + \pi\inv \zeta (3) \big[  g^{-2}_B +  \g_{3/2} 
    +  O(e^{-{2\pi \ov g_B}}) \big] +
 h_{1/2}  R\inv_B     (\nabla^2)^{1/2}_B
$$
\eqn\cono{
+\  
 \sum_{k=2}^\infty  
h_k \big[ \ 1\  +\  \g_{k-1/2}\ g^{2k-2}_B \  + \ 
 O(e^{-{2\pi \ov g_B}}) \big] \  R^{-2k}_B\   (\nabla^{2k})_B \bigg)\ 
( \RR^4)_B \ .  
}
The  terms proportional to 
$E_{k-1/2}(\Omega )$  thus  appear to contain 
only  one-loop and $k$-loop  parts  among   perturbative
 contributions.
 This  is 
a  generalisation of  the observation of  \refs{\grgu , \ggv}
about  $E_{3/2} (\O) \aa^{-1/2} \RR^4$ term 
(which contains tree-level and one-loop
contributions).
 It  seems likely that 
$O(e^{-{2\pi \ov g_B}})$  terms  in the expansion of  the functions
$E_{k-1/2}(\Omega )$ are
  related to  non-perturbative type II string  theory  contributions
since they  constitute the simplest $SL(2,{\bf Z})$ invariant completions
of the one-loop and $k$-loop terms.

\newsec{Acknowledgements}
We  
would like to thank 
 M.B. Green 
for useful    remarks.
 We   
 acknowledge
 the support of PPARC and 
 the European
Commission TMR programme grants  ERBFMRX-CT96-0045 and ERBFMBI-CT96-0982.
\vfill\eject
\listrefs
\end